\documentclass[11pt]{article}
\usepackage{moriond,epsfig}

\def\be{\begin{equation}}
\def\ee{\end{equation}}
\def\bea{\begin{eqnarray}}
\def\eea{\end{eqnarray}}

\def\lesssim{\mathrel{\hbox{\rlap{\hbox{\lower4pt\hbox{$\sim$}}}\hbox{$<$}}}}
\def\gtrsim{\mathrel{\hbox{\rlap{\hbox{\lower4pt\hbox{$\sim$}}}\hbox{$>$}}}}

\begin{document}
\vspace*{4cm}
\title{ON THE NEUTRINO FLUX FROM GAMMA RAY BURSTS: AN OVERVIEW}

\author{DAFNE GUETTA}

\address{Osservatorio astrofisico di Arcetri, Largo E. Fermi 5, \\
50100 Firenze, Italy}

\maketitle\abstracts{
Observations imply that gamma-ray bursts (GRBs) are produced by the
dissipation of the kinetic energy of a highly relativistic fireball.
Photo-meson interactions of protons with $\gamma$-rays
within the fireball dissipation region are expected to convert 
a significant fraction of the fireball energy into high energy neutrinos.
In this talk we summarize the results of an analysis of 
the internal shock model of GRBs, where
production of synchrotron photons and photo-meson neutrinos
with energies  $\sim 10^{14}$ eV are self-consistently calculated.
These neutrinos will be coincident with the GRB.
We show that the fraction of fireball energy converted 
into high energy neutrinos is not sensitive to uncertainties
in the fireball model parameters, such as the expansion Lorentz factor
and characteristic variability time.
Other processes of neutrinos emission from GRBs are also reviewed.
Photomeson interactions within the plasma region shocked by the
reverse shock, may produce a burst of $\sim 10^{18}$ eV neutrinos
following the GRB on the time scale of 10 s. Inelastic p-n 
nuclear collisions result in the production of a burst of 
$\sim$ 10 GeV neutrinos in coincidence with the GRB.
Planned 1 km$^3$ neutrino telescopes are expected to detect
ten 100 TeV neutrino events and several $10^{18}$ eV events,
correlated with GRBs, per year. A suitably densely spaced detector 
may allow the detection of several 10 GeV events per year.}

\section{Introduction}
The characteristics of $\gamma$-ray bursts (GRBs),
bursts of 0.1 MeV--1 MeV photons lasting for a few seconds
\cite{Fishman}, suggest that the observed radiation is produced by 
the dissipation of the kinetic energy
of a relativistically expanding wind, a 
``fireball,'' at cosmological distance
(see, e.g., \cite{fireballs} for review). 
The recent detection of delayed low energy (X-ray to radio) emission 
(afterglow) from GRB sources (see \cite{AG_ex_review} for review),
confirmed both the cosmological origin of the bursts and 
the standard model predictions for afterglows,
that result from the collision of an expanding fireball with
its surrounding medium (see \cite{AG_th_review} for review). 

Within the fireball model framework, observed $\gamma$-rays are produced
by synchrotron emission of electrons accelerated 
in the fireball dissipation region. In this region
the plasma parameters  
allow Fermi shock acceleration of protons up to energies 
$>10^{20}$~eV \cite{W95a,Vietri95};
(see \cite{My_revs} for a recent review).

High energy protons may photo-interact with the fireball photons
leading to emission of high energy neutrinos \cite{WnB97,WnB00}.
Electrons and protons can be accelerated to high energy by
internal shocks within the expanding wind or by the reverse
shock during the transition of the plasma to self-similarity.
In the first case
a burst of $\sim$ 100 TeV neutrinos 
is expected in coincidence with the GRB, and if protons are 
accelerated in the reverse shock
a burst of $\sim 10^{18}$ eV neutrinos is expected to follow the GRB
on a time scale of 10 s. Lower energy neutrinos
may be produced by inelastic nuclear collisions \cite{BnM00,MnR00}.

In this review new results found in Guetta, Spada \& Waxman 2001
(GSW01) \cite{guetta2} on the $\nu$ flux expected in 
the internal shock model scenario  are presented and
recent work on the high energy 
neutrino production by GRB fireballs is summarized.
Implications for planned $\sim$ 1${\rm km}^3$ neutrino
telescopes (the ICECUBE extension of AMANDA, ANTARES, 
NESTOR; see \cite{Alvarez00} for review) are discussed.  
High energy neutrino production in internal shocks is
discussed in \S~\ref{sec:ISnu}, other mechanisms for 
neutrino production in GRBs are summarized in 
\S~\ref{sec::other} and the implications for the planned 
neutrino telescopes are presented
in \S~\ref{sec::Implications}.

\section{Internal shocks neutrinos}
\label{sec:ISnu}
High energy neutrinos are produced by the decay of charged pions,
$\pi^+\rightarrow\mu^+\ +\ \nu_{\mu}\rightarrow
e^+ \ +\ \nu_e\ +\ \bar{\nu}_\mu \ +\nu_{\mu},$
created in interactions between the fireball photons and accelerated 
protons \cite{WnB97,WnB99}. 
The characteristic energy of these neutrinos  
is determined by the relation  between the observed photon energy, $E_\gamma$,
and the accelerated proton's energy, $E_p$,
at the photo-meson threshold  of the $\Delta$-resonance.
In the observer frame,
\begin{equation}
E_\gamma \, E_{p} = 0.2 \, {\rm GeV^2} \, \Gamma^2.
\label{eq:keyrelation}
\end{equation}
For typical observed $\gamma$-ray energy of 1~MeV
and Lorentz factors of the
expanding fireball $\Gamma \sim 300,$
proton energies $E_p\approx 2\times10^7$~GeV are 
required to produce neutrinos
from pion decay. Typically the neutrinos receive 5\% of proton
energy leading to $\nu$ of $\sim 10^{15}$ eV.

The predicted flux of $\gtrsim10^{14}$~eV neutrinos 
produced in internal shocks 
is determined by the fraction $f_\pi$ of fireball 
proton energy lost to pion production.
This fraction  is given by 
$f_{\pi}={\rm min}(1,\Delta t/t_{\pi})$ where $\Delta t$ is
the comoving shell expansion time and $t_{\pi}$ is the proton 
photo-pion energy loss time.
The value of this fraction has been estimated to be  \cite{WnB97,WnB99}
\begin{equation}
f_\pi(E_p)\approx0.2\min(1,E_p/E_p^b){L_{\gamma,52}\over
\Gamma_{2.5}^4 \Delta t_{-2}E_{\gamma,\rm MeV}^b}.
\label{eq:fpi}
\end{equation}
The $\gamma$-ray luminosity $L_\gamma$, 
the photon spectral break energy $E_\gamma^b$ (where 
luminosity per logarithmic photon energy interval peaks),
the observed variability time $\Delta t$
and the wind Lorenz factor $\Gamma$ 
are normalized in Eq. \ref{eq:fpi} to their
typical values inferred from observations:
$L_{\gamma}=10^{52}{\rm erg/s}$, $E_{\gamma}^b=1E_{\gamma,\rm MeV}^b$~MeV, 
$\Delta t=10^{-2}$s,
and $\Gamma=10^{2.5}$. The proton break energy, $E_p^b$, is the threshold
proton energy for interaction with photons of observed energy
$E_{\gamma}^b$,
$E_p^b=(2/E_{\gamma,\rm MeV}^b)\Gamma_{2.5}^2\times10^7$~GeV.
For a typical burst, $E\approx 10^{53}$ erg at redshift $z \sim 1$,
Eq. \ref{eq:fpi} implies a neutrino fluence
\begin{equation}
E_\nu^2\Phi_{\nu}\approx
10^{-3}\left(\frac{E_\nu}{E^b_\nu}\right)^{\alpha}\frac{\rm GeV}{\rm cm^2},
\label{eq:JGRB}
\end{equation}
where the neutrino break
energy is given by
$E_\nu^b\approx10^{15}{\Gamma_{2.5}^2/E_{\gamma,\rm MeV}^b}{\rm eV},$
whereas $\alpha=0$ for $E_\nu>E_\nu^b$ and $\alpha=1$ for $E_\nu<E_\nu^b$. 

The value of $f_\pi$, Eq. \ref{eq:fpi}, is strongly dependent on 
$\Gamma$. It has recently been pointed out by Halzen
and Hooper \cite{halzen} that if the Lorentz 
factor $\Gamma$ varies significantly between bursts, the resulting neutrino
flux will be dominated by a few neutrino bright bursts, and may 
significantly exceed 
the flux given by Eq. \ref{eq:JGRB}, derived 
for typical burst parameters.
This may strongly enhance the detectability of GRB neutrinos
by planned neutrino telescopes \cite{Alvarez00}.

In this section we present the most important results of 
the work GSW01 \cite{guetta2} whose main goal has been 
to determine the allowed range of variation in the 
fraction of fireball energy converted to high energy neutrinos, under the 
assumption that GRBs are produced
by internal dissipative shocks in an ultra-relativistic wind.

\subsection{Model dynamics and neutrino production}\label{subsec:prod}
In the fireball model of GRB \cite{fireballs} a compact source, of linear scale
$R_0\sim10^6$~cm, produces a wind characterized by an average luminosity 
$L_w\sim10^{52}{\rm erg\,s}^{-1}$--$10^{53}{\rm erg\,s}^{-1}$
and mass loss rate $\dot M=L_w/\eta c^2.$
Variability of the source on time scale $\Delta t$, resulting
in fluctuations in the wind bulk Lorentz factor $\Gamma$ on a similar
time scale, leads to internal shocks
that can reconvert a substantial 
part of the kinetic energy to internal energy. It is assumed that
this energy is then radiated as 
$\gamma$-rays by synchrotron (and inverse-Compton) emission of
shock-accelerated electrons.

In GSW01 we have used an approximate model for the
unsteady wind following 
Spada, Panaitescu \& M\'esz\'aros 2000 \cite{SPM}, 
and Guetta, Spada \& Waxman 2000
(GSW00) \cite{guetta1}. The wind flow is approximated as a 
set of discrete shells. 
Each shell is characterized by four 
parameters: the ejection time $t_j$, where the subscript $j$ denotes the
$j$-th shell, the Lorentz factor $\Gamma_j$, the mass $M_j$, 
and the width $\Delta_j$. 
Since the wind duration, $t_w\sim10$~s, is much larger than the dynamical
time of the source, $t_d\sim R_0/c$, variability of the wind on a wide range
of time scales, $t_d<t_v<t_w$, is possible. For simplicity, we 
considered a case where the wind variability is 
characterized by a single time scale 
$t_v$, in addition to the dynamical time scale of the source $t_d$
and to the wind duration $t_w$.
Thus, we considered shells of initial thickness 
$\Delta_j=c t_d=R_0$, ejected from the source
at an average rate $t_v^{-1}$. 
In GSW00 we  examined the dependence of the observed $\gamma$-ray 
flux and spectrum on wind model parameters, taking into account
both synchrotron and inverse-Compton emission, 
and the effect of $e^\pm$ pair production. 
We have shown that in order to obtain $\gamma$-ray fluxes and spectra 
consistent with observations, large variance is required in the 
wind Lorentz
factor distribution. We therefore  restricted our study to 
the bimodal case, where Lorentz factors are drawn from a bimodal distribution,
$\Gamma_j=\Gamma_m$ or $\Gamma_j=\Gamma_{M}\approx\eta_*\gg\Gamma_m$, 
with equal probability.
The time intervals $t_{j+1}-t_j$ were drawn randomly from
a uniform distribution with an average value of $t_v$.
In GSW00 we  considered two qualitatively 
different scenarios for shell masses distribution,
shells of either equal mass or equal energy, and concluded that
observations favor shells of equal mass. We therefore restricted
our analysis to the equal  shell mass case.

Once the shell parameters had been determined, we calculated the radii where  
collisions occur and determined the photon and neutrino emission 
from each collision. 
In each collision a forward and a reverse shock are formed,
propagating into forward and backward shells respectively.
We assumed that a fraction $\epsilon_e$  of the protons
shock thermal energy is converted to electrons and a fractio
$\epsilon_B$ to magnetic field.
We  adopted  these fractions to be close to equipartition, 
$\epsilon_e=0.45$ and $\epsilon_B\gtrsim 0.01$.
We further  assumed 
that both electrons and protons are accelerated by 
the shocks to a a power-law distribution, 
$dn_\alpha/d\gamma_\alpha\propto\gamma_\alpha^{-p},$ with $p\approx 2$ 
for particle
Lorentz factors $\gamma_{\alpha,\min}<\gamma_\alpha<\gamma_{\alpha,\max}$. 
The maximum Lorenz factor is determined by equating the acceleration
time, estimated as the Larmor radius divided
by $c$, to the minimum between the dynamical  and  synchrotron 
cooling time.

The calculation of the emitted $\nu$ spectrum and flux was
carried out according to the method described in GSW01, 
following Waxman \& Bahcall 1997 \cite{WnB97}. 
Photo-meson production of low energy protons, well below
$E_p^b$, requires interaction with 
high energy photons, well above $E_\gamma^b$. Such photons
may be depleted by pair production. 
For each collision we found the photon energy 
$E_{\gamma}^{\pm}$, for which the pair production optical thickness,
$\tau_{\gamma\gamma}$, is unity. A large fraction of 
photons of energy exceeding $\max(E^{\pm}_{\gamma},m_e c^2)$ (measured
in the shell frame) will be converted
to pairs, and hence will not be available for photo-meson interaction,
leading to a suppression of the neutrino flux
at low energies. In order to take this effect into account,
we  used  
$f_{\pi}={\rm min}[1,(\Delta t/t_{\pi})\min(1,\tau_{\gamma\gamma}^{-1})]$
for protons interacting with photons which are strongly suppressed by pair
production.

Neutrino production is suppressed at high energy,
where neutrinos are produced by the decay of muons and pions 
whose lifetime $\tau$ exceeds the characteristic 
time for energy loss due to synchrotron emission 
\cite{WnB97,WnB99,rachen}.
We therefore  defined an effective $f_\pi$, $f_{\pi,\rm eff.}$, as 
$4f_\pi$ times the fraction of pions' energy converted to muon neutrinos.

In each collision a fraction 
of the kinetic energy of the colliding shell  is converted to a flux 
of photons and neutrinos. The energy that is not lost to 
photons and neutrinos is converted back to kinetic energy by adiabatic
expansion of the shell.

\subsection{Main results and conclusions}
\label{sec::results}
\begin{figure}
\psfig{figure=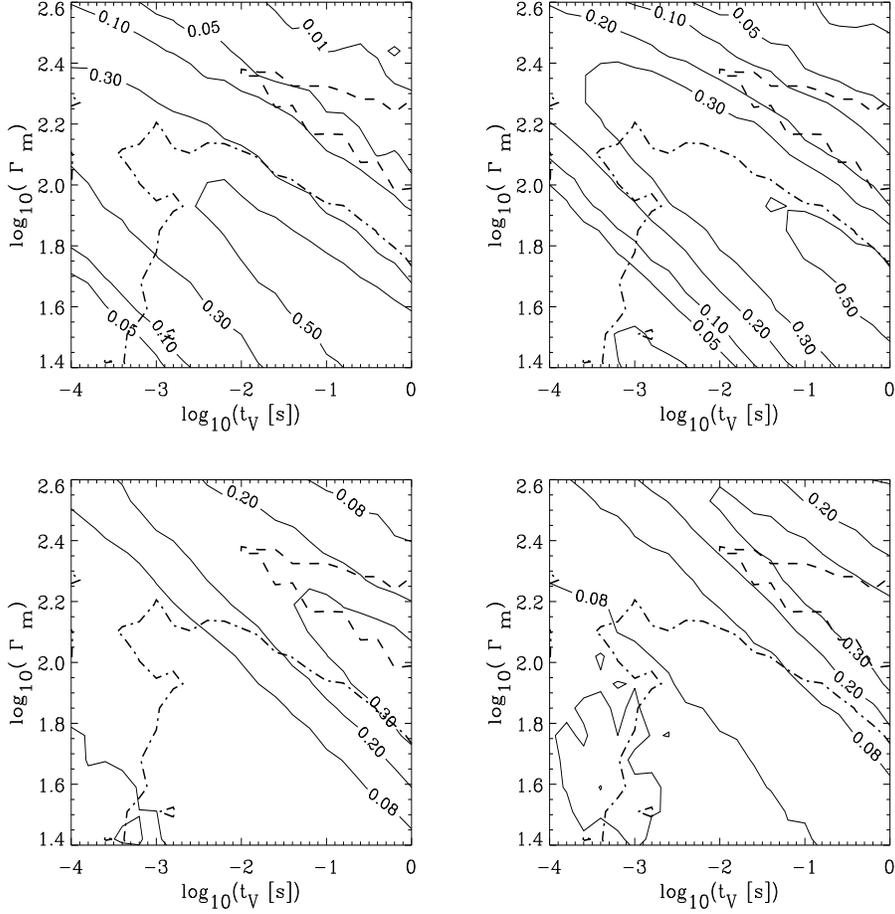,height=5in,width=5in}
\caption{Contour plots
of $f_{\pi,\rm eff.}$, the effective value of $f_\pi$ 
(defined as $4f_\pi$ times the fraction of pions' 
energy converted to muon neutrinos), as function of 
wind variability time $t_v$ and minimum Lorentz factor $\Gamma_m$, 
for wind luminosity $L_w=10^{53}{\rm erg/s}$ and $\epsilon_B=0.01$. 
The four panels correspond to four observed neutrino
energy bins, clockwise from top left:  
$10^{14}{\rm eV}<E_{\nu}<10^{15}{\rm eV}$,
$10^{15}{\rm eV}<E_{\nu}<10^{16}{\rm eV}$,
$10^{17}{\rm eV}<E_{\nu}<10^{18}{\rm eV}$,
$10^{18}{\rm eV}<E_{\nu}<10^{19}{\rm eV}$. We have used the approximate
relation $E_\nu=0.05E_p$ between neutrino and proton energy, and assumed
a source redshift $z=1.5$. The region in the 
$\Gamma_m$--$t_v$ plane where $E_\gamma^b>0.1$~MeV
is bound by the dashed lines.
The dash-dotted lines
outline the region in which 
the fraction of wind energy converted to radiation
exceeds 2\% (a higher fraction is obtained at larger $\Gamma_m$ values).
\label{fig:1}}
\end{figure}
\begin{figure}
\psfig{figure=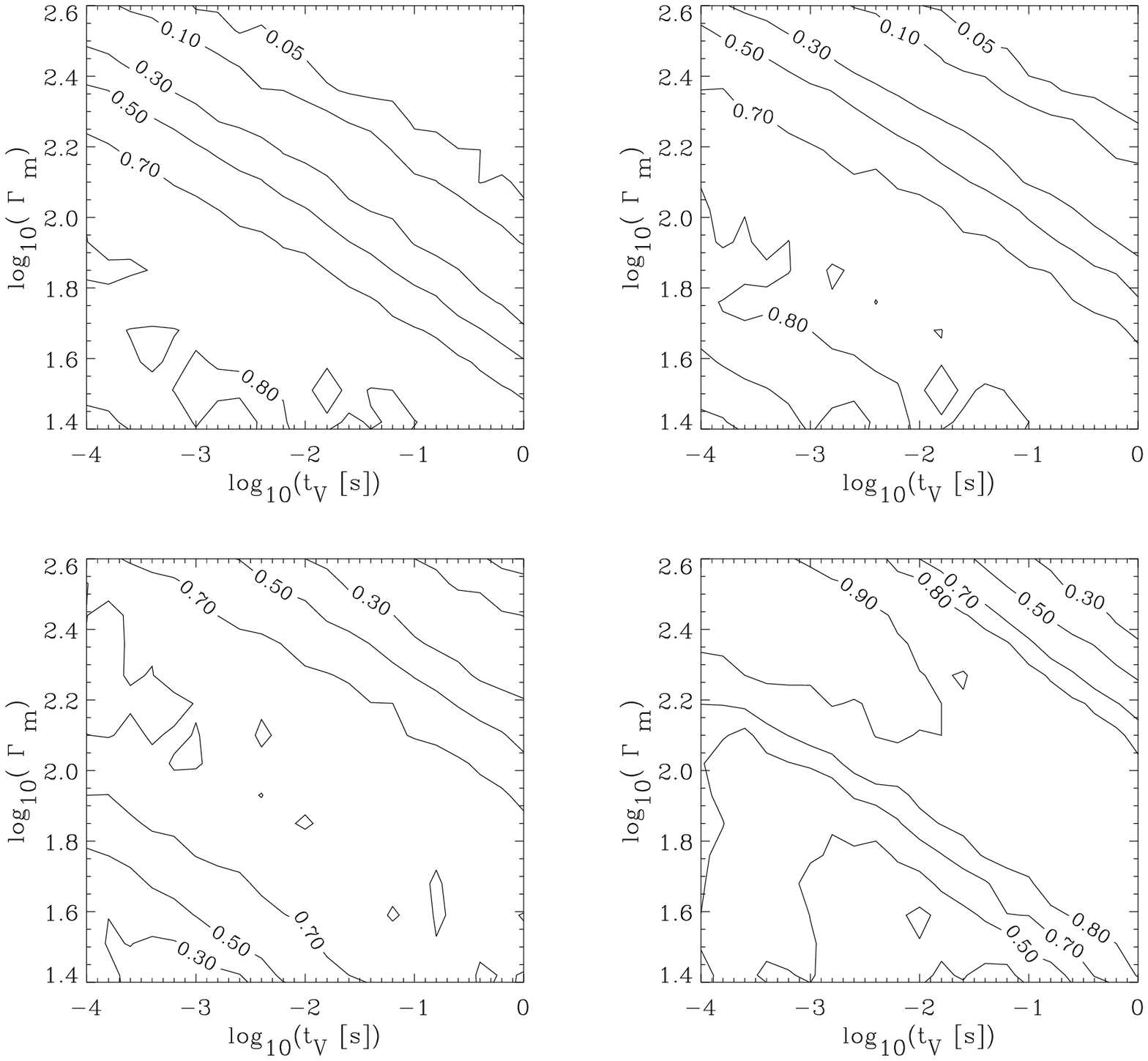,height=5in,width=5in}
\caption{Contour plots
of $f_\pi$, obtained when
the effects on neutrino production of synchrotron losses and
pair-production are neglected, for the case shown in Figure 1.
\label{fig:2}}
\end{figure}
The main results of GWS01 are presented in Figures 1 and 2.
Figure 1 presents the contour plot
for the effective $f_\pi$, $f_{\pi,\rm eff.}$
taking $\epsilon_B=0.01$.
The values shown are averaged over all internal collisions, 
weighted by the energy converted to high energy protons in each collision. 
Values of $f_\pi$ are shown for several neutrino energy ranges, using
the approximation $E_\nu=0.05E_p$. The region in 
$\Gamma_m$--$t_v$ plane where $E_\gamma^b>0.1$~MeV
is bound in Figures 1 by the dashed line.
The dash-dotted line outlines the region in which 
the fraction of wind energy converted to radiation
exceeds 2\% (higher fraction is obtained for larger $\Gamma_m$ values).
Figure 2 shows the contour plots of $f_{\pi}$, neglecting the 
effects on neutrino production of synchrotron losses and pair-production.
$f_{\pi}$ approaches unity at low values of $\Gamma_m$ for low
proton (neutrino) energies, and at intermediate values of $\Gamma_m$
for high proton (neutrino) energies. 
The energy loss of pions and
muons due to synchrotron emission reduces the fraction of pion energy
converted to neutrinos, and hence suppresses the effective value of
$f_{\pi}$, to the range of values shown in Figure 1. The suppression
seen in Figure 2
of $f_\pi$ at low $\Gamma_m$ for high energy protons is due to the fact
that for low values of $\Gamma_m$ most collisions occur at small radii, 
where synchrotron losses of high energy protons prevent acceleration of 
protons to ultra-high energy \cite{W95a}. Ultra-high energy protons are
produced in this case only by secondary collisions at large radii, where
the photon energy density is already low, resulting in smaller values of
$f_\pi$.

As we can see from Figure 1, for wind parameters
consistent with observed GRB characteristics,
the effective value of $f_\pi$ at high neutrino energies, $E_\nu>E_\nu^b$,
is restricted to values in the range of $\approx10\%$ to 
$\approx30\%$. The weak dependence of
$f_{\pi,\rm eff.}$ on wind the model parameters, in contrast with the strong
dependence implied by Eq. \ref{eq:fpi}, is
due to two reasons.
First, for low values of $\Gamma$ and $\Delta t$, where large values 
of $f_\pi$ are  implied by Eq. \ref{eq:fpi}, 
only a small fraction of the pions' energy is converted
to neutrinos at high proton energy due to pion and muon synchrotron losses
(compare figures 1 and 2). Secondth, 
the observational constraints imposed by $\gamma$-ray observations
imply that the wind model
parameters ($\Gamma$, $L$, $\Delta t$) are correlated.
From Figure 1 we note, also,  that 
the value of $f_{\pi,\rm eff.}$ does not significantly exceed 
$20\%$ also in wind model parameter regions which are outside the
parameter region implied by observations.
Our results imply that GRB neutrino flux of individual bursts
should correlate mainly with the bursts' $\gamma$-ray flux.

\section{Neutrinos from other processes in GRBs}
\label{sec::other}
\subsection{Neutrinos from the reverse shock}
\label{sec::afterglow}
Much higher energy, $\gtrsim 10^{18}$~eV, neutrinos may be produced at a later
stage, at the onset of the fireball interaction with its surrounding medium. 
During the transition to self-similar expansion, which takes 
place on time scale $\sim t_w,$ protons and electrons are
accelerated in the reverse shocks.
The combination of low energy photons, 10 eV-1 keV,
and high energy protons produces
neutrinos of ultra-high energy $10^{17}-10^{19}$ eV
via photo-meson interactions, as indicated by
Eq. \ref{eq:keyrelation} \cite{WnB00}.
However, the fraction of energy converted to pions
depends upon parameters of the surrounding medium.
If the density of the surrounding gas is that 
typical of the interstellar medium 
$n_0\sim 1 {\rm cm}^{-3},$ the photon emission peaks in the 
X-ray band, $E_{\gamma,\rm keV}^b$ and the luminosity is
$L_{\rm X}\approx 10^{50}{\rm erg/s}.$
Using these parameters in Eq. \ref{eq:keyrelation},
replacing $\Delta t$ with $t_w$ and 
considering  that the self-similar Lorentz factor
of the expanding plasma 
under these conditions is $\Gamma_{\rm BM}\sim 250,$
we find $f_{\pi}=0.01.$
Thus the expected neutrino fluence for a typical burst,
$E\approx 10^{53}$ erg at $z\sim 1$, is \cite{WnB00}
\begin{equation}
E_\nu^2\Phi_{\nu}\approx
10^{-4.5}
\left(\frac{E_\nu}{E^b_\nu}\right)^{\alpha} \frac{{\rm GeV}}{{\rm cm^2}},
\label{eq:J2GRB}
\end{equation}
where $E^b_\nu\sim 10^{17}$ eV, $\alpha=1/2$ for 
$E_\nu<E^b_\nu$ and $\alpha=1$ for $E_\nu>E^b_\nu.$
If the fireball expands into a pre-existing wind, 
the transition to self-similar behaviour takes place at
a radius where the wind density is $n\approx 10^4 {\rm cm}^{-3}.$
Since  $\Gamma_{\rm BM}$ decreases with increasing density,
according to Eq. \ref{eq:keyrelation} protons of energy 
$E_p>10^{18}$ eV lose all their energy to pion production in 
the wind case, and a typical GRB at $z\sim 1$ is expected to
produce a neutrino fluence \cite{WnB00}
\begin{equation}
E_\nu^2\Phi_{\nu}\approx
10^{-2.5}
\left(\frac{E_\nu}{E^b_\nu}\right)^{\alpha} \frac{{\rm GeV}}{{\rm cm^2}},
\label{eq:J3GRB}
\end{equation}
where again  $E^b_\nu\sim 10^{17}$ eV, $\alpha=0$ for 
$E_\nu<E^b_\nu$ and $\alpha=1$ for $E_\nu>E^b_\nu.$

\subsection{Inelastic p-n collisions}
We briefly discuss how lower energy neutrinos may be 
produced by inelastic nuclear collisions, 
a detailed analysis of these processes is given in  \cite{MnR00,BnM00}.
According to most progenitor scenarios neutrons can contribute
to the barionic component of the fireball as much as protons.
During the acceleration phase of the fireball 
protons and neutrons are coupled by nuclear scattering. 
As the fireball expands neutrons and protons may decouple and relativistic 
relative velocities may arise between them, leading 
to pion production through unelastic collisions.
Since the decuopling happens when the wind
Lorentz factor is $\sim 400,$ neutrons decouple
from the accelerating plasma prior to saturation,
$\Gamma=\eta$, only if $\eta>400.$
Charged pions may decay in $\sim$ 10 GeV neutrinos.
A typical burst, $E=10^{53}$ erg at $z\sim 1,$ with a significant
neutron to proton ratio and $\eta>400$ will produce a fluence of 
$\sim 10^{-4}{\rm cm}^{-2}$ of $\sim 10$ GeV neutrinos. 

\section{Implications}
\label{sec::Implications}
The high energy neutrinos predicted in the dissipative wind 
model of GRBs may be observed by detecting the Cerenkov light 
emitted by high energy muons produced by neutrino 
interactions below a detector on the surface of the Earth
(see \cite{gaisser} for a recent review).

For the internal shocks neutrinos with energies $10^{14}-10^{15}$
eV, the fluence found in Eq. \ref{eq:JGRB} implies a detection
probability of $\sim 0.03$ per burst in a km-cube telescope,
corresponding to tens of events per years correlated in time 
and direction with GRBs, given the observed GRB rate of 
$\approx 10^3 {\rm yr}^{-1}.$ 
The predicted fluence of neutrinos of $10^{17}$ eV
produced during the transition of the fireball to self-similarity
depends strongly upon the conditions of the surrounding gas.
In the case of a fireball expanding in a typical
interstellar medium Eq. \ref{eq:J2GRB} implies a detection
probability of $\sim 10^{-4.5}$ per burst in a km-cube telescope.
This probability is higher, $\sim 10^{-2.5},$ in the case of 
fireball expansion into a pre-existing massive star wind.
In this case a km$^3$ telescope  should detect several 
muon induced neutrinos per year.

Inelastic p-n collisions may produce neutrinos of $\sim$ 10 GeV 
with a fluence of $\sim 10^{-4}$ cm $^{-2}$ per burst, due 
to the decoupling of neutrons and protons during the 
acceleration phase.  
This process is possible only if large neutron to proton
ratio and high, $>400$, Lorentz factor are assumed.
Ten events per year are expected in a km$^3$ detector.
Such events may be detectable in a suitably densely spaced detector.

Detection of high energy neutrinos will test the shock acceleration
mechanism and the suggestion that GRBs are the sources of ultra-high
energy protons, since $\ge 10^{14}$ eV ($\ge 10^{18}$ eV) neutrino
production requires protons of energy $\ge 10^{16}$ eV 
($\ge 10^{19}$ eV).
Detection of $\sim$ 10 GeV neutrinos will constrain the fireball 
neutron fraction, and hence the GRB progenitor. 

\section*{Acknowledgments}
I thank Eli Waxman for his fundamental help in preparing this review
and the original work contained in it.

\end{document}